\documentclass[a4paper,twocolumn,floatfix,amsmath,amssymb,fleqn,prb,showkeys,showpacs]{revtex4}

\usepackage{graphicx}
\usepackage{dcolumn}
\usepackage{bm}
\usepackage{color}
\begin{document}
\newcommand{\newc}{\newcommand}
\newc{\mbf}{\mathbf}
\newc{\boma}{\boldmath}
\newc{\beq}{\begin{equation}}
\newc{\eeq}{\end{equation}}
\newc{\beqar}{\begin{eqnarray}}
\newc{\eeqar}{\end{eqnarray}}
\newc{\beqa}{\begin{eqnarray*}}
\newc{\eeqa}{\end{eqnarray*}}

\newc{\bd}{\begin{displaymath}}
\newc{\ed}{\end{displaymath}}

\title{On Virtual Displacement and Virtual Work in Lagrangian Dynamics}

\author{Subhankar Ray}
\email{sray_ju@rediffmail.com, subho@juphys.ernet.in (S. Ray)} 
\affiliation{Department of Physics, Jadavpur University, Calcutta 
700 032, India}
\affiliation{C. N. Yang Institute for Theoretical Physics, 
Stony Brook, NY 11794}
\author{J. Shamanna}
\email{jaya@vbphysics.net.in  (J. Shamanna)}
\affiliation{Physics Department, Visva Bharati University, 
Santiniketan 731235, India}

\date{August 1, 2005}

\begin{abstract}
The confusion and ambiguity encountered by students, in 
understanding virtual displacement and virtual work, is
discussed in this article.
A definition of virtual displacement is presented that allows
one to express them explicitly for holonomic (velocity independent),
non-holonomic (velocity dependent), scleronomous (time independent)
and rheonomous (time dependent) constraints.
It is observed that for holonomic, scleronomous constraints,
the virtual displacements are the displacements allowed by the constraints.
However, this is not so for a general class of constraints. For simple 
physical systems, it is shown that, the work done by the constraint forces 
on virtual displacements is zero. 
This motivates Lagrange's extension of d'Alembert's 
principle to system of particles in constrained motion.
However a similar zero work principle does not hold for the allowed displacements.
It is also demonstrated that d'Alembert's principle of zero virtual work
is necessary for the solvability of a constrained mechanical problem.
We identify this special class of constraints, physically
realized and solvable, as {\it the ideal constraints}.
The concept of virtual displacement and the principle
of zero virtual work by constraint forces are central to both
Lagrange's method of undetermined multipliers, and Lagrange's
equations in generalized coordinates.
\end{abstract}

\maketitle

\section{Introduction}
Almost all graduate level courses in classical mechanics
include a discussion of virtual displacement 
\cite{goldstein,sommer,hylleraas,greenwood,schaum,symon,sygr,
taylor,haas,terhaar,hand} and
Lagrangian dynamics 
\cite{goldstein,sommer,hylleraas,greenwood,schaum,symon,sygr,
taylor,haas,terhaar,hand,landau}. From the concept 
of zero work by constraint forces on virtual 
displacement, the Lagrange's equations of motion are derived. 

However, the definition presented in most accessible texts 
often seem vague and ambiguous to students.
Even after studying the so called definition, it is rather
commonplace that a student fails to identify,
whether a supplied vector is suitable as a virtual displacement,
for a given constrained system.
Though some of the more advanced and rigorous treatise
\cite{arnold,pars} present a more precise and satisfactory 
treatment, they are often not easily comprehensible to most
students.
In this article we attempt a simple, systematic and precise 
definition of virtual displacement, which clearly shows the
connection between the constraints and the corresponding allowed
and virtual displacements. This definition allows one to understand 
how far the virtual displacement is `arbitrary' and how far it 
is `restricted' by the constraint condition.

There are two common logical pathways of arriving at Lagrange's equation.
\begin{enumerate}
\item Bernoulli's principle of virtual velocity \cite{sommer} (1717),
d'Alembert's principle of zero virtual work \cite{sommer,dalembert} (1743),
Lagrange's generalization of d'Alembert's principle to constrained system
of moving particles, and Lagrange's equations of motion (1788)
\cite{sommer,lagr1,lagr2}.
\item Hamilton's principle of least action \cite{sommer,hamilton}
(1834), and variational approach to Lagrange's equation.
\end{enumerate}
The two methods are logically and mathematically independent and 
individually self contained. 
The first method was historically proposed half a century
earlier, and it presents the motivation of introducing the
Lagrangian as a {\it new physical quantity}.
The second method starts with the Lagrangian and the related action
as quantities axiomatically describing the dynamics of the system. 
This method is applied without ambiguity in some
texts\cite{landau,arnold} and courses \cite{drell}.
However one also finds intermixing of the two approaches in 
the literature and popular texts,
often leading to circular definition and related confusion. 
A rational treatment demands an independent presentation of the two
methods, and then a demonstration of their interconnection.
In the present article we confine ourselves to the first method.

In this approach, due to Bernoulli, d'Alembert and Lagrange, one begins 
with a constrained system, defined by equations of constraints connecting
positions, time and often velocities of the particles under consideration.
The concept of virtual displacement is introduced in terms of the 
constraint equations. The external forces alone cannot maintain the 
constrained motion. This requires the introduction of forces of
constraints. The imposition of the principle of zero virtual work by
constraint forces gives us a `special class of systems', that are
solvable.

A proper definition of virtual displacement is necessary to
make the said approach logically satisfactory. However the various
definitions found in popular texts are often incomplete and
contradictory with one another.
These ambiguities will be discussed in detail in the next section.

In the literature, e.g., Greenwood \cite{greenwood} Eq.1.26 and 
Pars \cite{pars} Eq.1.6.1, one encounters holonomic constraints 
of the form:
\beq\label{const_holo}
\phi_j(x_1,x_2,\dots,x_{3N},t) = 0\,, \hfill j=1,2,\dots,k.
\eeq
The differential form of the above equations are satisfied by 
allowed infinitesimal displacements $\{dx_i\}$,
(Greenwood\cite{greenwood} Eq.1.27; Pars\cite{pars} Eq.1.6.3).
\beq\label{adisp}
\sum_{i=1}^{3N} \frac{\partial \phi_j}{\partial x_i} dx_i
+ \frac{\partial \phi_j}{\partial t} dt = 0 \,, \hfill j=1,2,\dots,k.
\eeq
For a system under above constraints 
the virtual displacements $\{\delta x_i\}$, satisfy the following 
equations (Greenwood\cite{greenwood} Eq.1.28; Pars\cite{pars} Eq.1.6.5),
\beq\label{vdisp}
\sum_{i=1}^{3N} \frac{\partial \phi_j}{\partial x_i} \delta x_i
= 0\,, \hfill j=1,2,\dots,k.
\eeq
The differential equations satisfied by allowed and
virtual displacements are different even for the non-holonomic case.
Here, the equations satisfied by the allowed displacements $\{dx_i\}$ are
(Goldstein \cite{goldstein} Eq.2.20, Greenwood\cite{greenwood} 
Eq.1.29 and Pars\cite{pars} Eq.1.7.1),
\beq\label{adisp_nh}
\sum_{i=1}^{3N} a_{ji} dx_i + a_{jt} dt = 0\,,  \hfill
j=1,2,\dots,m.
\eeq
Whereas, the virtual displacements $\{\delta x_i\}$
satisfy (Goldstein\cite{goldstein} Eq.2.21, Greenwood\cite{greenwood}
Eq.1.30; Pars\cite{pars} Eq.1.7.2),
\beq\label{vdisp_nh}
\sum_{i=1}^{3N} a_{ji} \delta x_i = 0\,, \hfill j=1,2,\dots,m.
\eeq
Thus, there appear in the literature certain equations, namely
Eq.(\ref{vdisp}) and Eq.(\ref{vdisp_nh}),
which are always satisfied by the {\it not so
precisely defined} virtual displacements. It may be noted that these
equations are connected to the constraints but are not simply the
infinitesimal forms of the constraint equations, i.e.,
Eqs.(\ref{adisp}) and (\ref{adisp_nh}).
This fact is well documented in the literature
\cite{goldstein,greenwood,pars}. However, the nature of the difference 
between these sets of equations, i.e, Eqs.(\ref{vdisp}) and (\ref{vdisp_nh})
on one hand and Eqs.(\ref{adisp}) and (\ref{adisp_nh}) on the other,
and their underlying connection are not explained in most discussions.
One may consider Eq.(\ref{vdisp}) or Eq.(\ref{vdisp_nh}), as 
independent defining equation for virtual displacement.
But it remains unclear as to how,
the virtual displacements $\{\delta x_i\}$ defined by two different sets
of equations for the holonomic and the non-holonomic cases, viz.,
Eq.(\ref{vdisp}) and Eq.(\ref{vdisp_nh}), correspond to the same concept of
virtual displacement.

We try to give a physical connection between the definitions of allowed
and virtual displacements for any given set of constraints.
The proposed definition of virtual displacement (Sec.IIA) as difference of two
unequal allowed displacements (satisfying Eq.(\ref{adisp}) or 
Eq.(\ref{adisp_nh})) 
over the same time interval; automatically ensures that virtual displacements
satisfy Eq.(\ref{vdisp}) and Eq.(\ref{vdisp_nh}) for holonomic and
non-holonomic systems respectively.
We show that in a number of natural systems, e.g.,
pendulum with fixed or moving support, particle 
sliding along stationary or moving frictionless inclined plane,
the work done by the forces of constraint on virtual displacements
is zero. 
We also demonstrate that this condition is necessary for the
solvability of a constrained mechanical problem.
Such systems form an important class of natural systems. 
\subsection{Ambiguity in virtual displacement}
In the literature certain statements appear in reference to
virtual displacement, which seem confusing and mutually
inconsistent, particularly to a student.
In the following we present few such statements found in common
texts.
\begin{enumerate}
%
%
\item It is claimed that (i){\it a virtual displacement $\delta\mbf{r}$
is consistent with the forces and constraints imposed on the system at
a given instant $t$} \cite{goldstein}; 
(ii) {\it a virtual displacement is an arbitrary, instantaneous, 
infinitesimal change of position of the system compatible with the
conditions of constraint} \cite{sommer};
(iii) {\it virtual displacements are, by definition, arbitrary displacements
of the components of the system, satisfying the constraint} \cite{hylleraas};
(iv) {\it virtual displacement does not violate the constraints} \cite{taylor}; 
(v) {\it we define a virtual displacement as one which does not violate
the kinematic relations} \cite{terhaar};
(vi) {\it the virtual displacements obey the constraint on the motion}
\cite{hand}.
These statements imply that the virtual displacements satisfy the constraint 
conditions, i.e., the constraint equations.
However this is true only for holonomic, sclerenomous constraints.
We shall show that for non-holonomic constraints, or rheonomous constraints, 
e.g., a pendulum with moving support, this definition violates
the zero virtual work principle.
%
%
\item It is also stated that 
(i) {\it virtual displacements do not necessarily conform to the 
constraints} \cite{greenwood};
(ii) {\it the virtual displacements $\delta q$ have nothing to do with
actual motion. They are introduced, so to speak, as test quantities,
whose function it is to make the system reveal something about its
internal connections and about the forces acting on it} \cite{sommer};
(iii) {\it the word ``virtual'' is used to signify that the displacements
are arbitrary, in the sense that they need not correspond to any actual
motion executed by the system} \cite{hylleraas};
(iv) {\it it is not necessary that it} (virtual displacement) {\it 
represents any actual motion of the system} \cite{symon}; 
(v) {\it it is not intended to say that such a displacement} (virtual)
{\it occurs during the motion of the particle considered, or even that
it could occur} \cite{haas};
(vi) {\it virtual displacement is any arbitrary infinitesimal 
displacement not necessarily along the
constrained path} \cite{schaum}. 
From the above we understand that the virtual displacements do not 
necessarily satisfy the constraint equations, and they need not be the 
ones actually realized.
We shall see that these statements are consistent with physical 
situations, but they cannot serve as a satisfactory definition of virtual 
displacement.
Statements like: ``not necessarily conform to the constraints'' or 
``not necessarily along the constrained path'' only tell us what 
virtual displacement is not, they do not tell us what it really is.
Reader should note that there is a conflict between the 
statements quoted under items 1 and 2. 

Thus it is not clear from the above, 
whether the virtual displacements
satisfy the constraints, i.e., the constraint equations, or they do not.
%
%
\item It is also stated that (i){\it virtual displacement is to be
distinguished from an actual displacement of the system occurring in a 
time interval $dt$} \cite{goldstein};  
(ii) {\it it is an arbitrary, instantaneous, 
change of position of the system} \cite{sommer};
(iii) {\it virtual displacement $\delta\mbf{r}$ takes place without any 
passage of time} \cite{taylor}.
(iv) {\it virtual displacement has no connection with the time - in
contrast to a displacement which occurs during actual motion, and which
represents a portion of the actual path} \cite{haas};
(v) {\it one of the requirements on acceptable virtual displacement
is that the time is held fixed} \cite{hand}.
We even notice equation like : ``$\delta x_i = d x_i$ for $dt=0$'' \cite{taylor}.
The above statements are puzzling to a student.
If position is a continuous function of time, a change in position during 
zero time has to be zero. In other words, this definition implies that 
the virtual displacement cannot possibly be an infinitesimal (or 
differential) of any continuous function of time. 
In words of Arthur Haas: {\it since its} (virtual displacement)
{\it components are thus not functions of the time, we are not able to
regard them as differentials, as we do for the components of the element
of the actual path} \cite{haas}.
It will be shown later (Sec.II), that virtual displacement can be looked 
upon as a differential. 
It is indeed a differential change in position or
an infinitesimal displacement, consistent with
virtual velocity $\widetilde{\mbf{v}}_k(t)$, taken over a time 
interval $dt$ (see Eq.(\ref{veldiff})).
\item Virtual displacement is variously described as: {\it arbitrary},
{\it virtual}, and {\it imaginary} \cite{goldstein,sommer,
hylleraas,symon,schaum}. These adjectives make the definition more
mysterious to a student.
\end{enumerate}

Together with the above ambiguities, 
students are often unsure whether it is sufficient to discuss 
virtual displacement as an abstract concept or it is important 
to have a quantitative definition.
Some students appreciate that the virtual displacement as a vector
should not be ambiguous. The principle of zero virtual work is required
to derive Lagrange's equations. For a particle under constraint this 
means that the virtual displacement is always orthogonal to the 
force of constraint.

At this stage a student gets further puzzled. Should he take the
forces of constraint as supplied, and the principle of
zero virtual work as a definition of virtual displacement ? 
In that case the principle reduces merely to a definition of a new 
concept, namely virtual displacement. 
Or should the virtual displacement be defined
from the constraint conditions independently ?
The principle of zero virtual work may then be used to obtain
the forces of constraint. These forces of constraint ensure that 
the constraint condition is maintained throughout the motion.
Hence it is natural to expect that they should be connected to
and perhaps derivable from the constraint conditions.


\section{Virtual displacement and Forces of Constraint}
\subsection{Constraints and Virtual displacement}
Let us consider a system of constraints
that are expressible as equations involving positions and time. 
They represent some geometric restrictions (holonomic)
either independent of time (sclerenomous) or explicitly dependent 
on it (rheonomous). Hence for a system of $N$ particles moving in
three dimensions, a system of ($s$) holonomic, rheonomous 
constraints are represented by functions of $\{\mbf{r}_k\}$ and ($t$),
\beq{\label{holo}}
f_i(\mbf{r}_1,\mbf{r}_2,\dots,\mbf{r}_N,t) = 0,
\hfill i=1,2,\dots,s.
\eeq
The system may also be subjected to non-holonomic constraints
which are represented by equations connecting velocities $\{\mbf{v}_k\}$,
positions $\{\mbf{r}_k\}$ and time ($t$).
\beq{\label{nonholo}}
\sum_{k=1}^{N} \mbf{A}_{ik} \cdot \mbf{v}_k + 
A_{it} = 0\,, 
\hfill i=1,2,\dots,m,
\eeq
where $\{\mbf{A}_{ik}\}$ and $\{A_{it}\}$ are functions
of positions $\{\mbf{r}_1,\mbf{r}_2,\dots,\mbf{r}_N\}$ and time ($t$).
The equations for non-holonomic constraints impose restrictions
on possible or allowed velocity vectors
$\{\mbf{v}_1,\mbf{v}_2,\dots,\mbf{v}_N\}$,
for given positions $\{\mbf{r}_1,\mbf{r}_2,\dots,\mbf{r}_N\}$ and time ($t$).
The holonomic constraints given by Eq.(\ref{holo}), are
equivalent to the following equations imposing further restriction
on the possible or allowed velocities.
\beq{\label{velconst}}
\sum_{k=1}^{N} \left( \frac{\partial f_i}{\partial \mbf{r}_k} 
\right) \cdot \mbf{v}_k + \frac{\partial f_i}{\partial t} = 0,
\hfill i=1,2,\dots,s.
\eeq
For a system of $N$ particles under ($s$) holonomic and ($m$) non-holonomic
constraints, a set of vectors $\{\mbf{v}_1,\mbf{v}_2,\dots,\mbf{v}_N\}$
satisfying Eq.(\ref{nonholo}) and Eq.(\ref{velconst}) are called
allowed velocities.
It is worth noting at this stage that there are many, in fact 
infinitely many, allowed velocities, since we have imposed only
($s+m$) number of scalar constraints, Eq.(\ref{nonholo}) and
Eq.(\ref{velconst}), on ($3N$) scalar components 
of the allowed velocity vectors. 

At any given instant of time, the difference of any two
such non-identical allowed sets of velocities, independently 
satisfying the constraint conditions, are
called virtual velocities.
\bd
\widetilde{\mbf{v}}_k(t) = \mbf{v}_k(t) - \mbf{v}'_k(t) 
\hfill k = 1,2,\dots, N.
\ed
An infinitesimal displacement 
over time ($t, t+dt$), due to allowed velocities, will be called 
the {\it allowed infinitesimal displacement} or simply allowed
displacement. 
\beq{\label{alldisp}}
d\mbf{r}_k = \mbf{v}_k(t) dt,
\hfill k=1,2,\dots,N.
\eeq
Allowed displacements $\{d\mbf{r}_k\}$ together with differential of
time ($dt$) satisfy the infinitesimal form of the constraint equations.
From Eq.(\ref{velconst}) and Eq.(\ref{nonholo}), we obtain, 
\beq{\label{allcons1}}
\sum_{k=1}^{N} \left( \frac{\partial f_i}{\partial \mbf{r}_k} 
\right) \cdot d\mbf{r}_k + \frac{\partial f_i}{\partial t} dt = 0,
\hfill i=1,2,\dots,s,
\eeq
\vskip -.5cm
\beq{\label{allcons2}}
\sum_{k=1}^{N} \mbf{A}_{ik} \cdot d \mbf{r}_k + A_{it} dt = 0\,,  
\hfill i=1,2,\dots,m.
\eeq
As there are many independent sets of
allowed velocities, we have many allowed sets of
infinitesimal displacements. We propose to define 
{\it virtual displacement} as the difference 
between any two such (unequal) allowed displacements 
taken over the same time interval ($t, t+dt$),
\beq{\label{virdisp}}
\delta \mbf{r}_k = d\mbf{r}_k - d\mbf{r}'_k,
\hfill k=1,2,\dots,N.
\eeq
Thus virtual displacements are infinitesimal displacements
over time interval $dt$ due to virtual velocity $\widetilde{\mbf{v}}_k(t)$,
\beq{\label{veldiff}}
\delta \mbf{r}_k = \widetilde{\mbf{v}}_k(t) dt = (\mbf{v}_k(t) - 
\mbf{v}'_k(t)) dt,
\hfill k=1,2,\dots,N.
\eeq
This definition is motivated by the possibility of (i) identifying a 
special class of `{\it ideal constraints}' (Sec.IIC), and
(ii) verifying `{\it the principle of zero virtual work}' in common
physical examples (Sec.III).
It may be noted that, by this definition, virtual displacements
$\{\delta\mbf{r}_k\}$ are not instantaneous changes in position in zero time.
They are rather smooth, differentiable objects.

\begin{figure}[h]
{\includegraphics{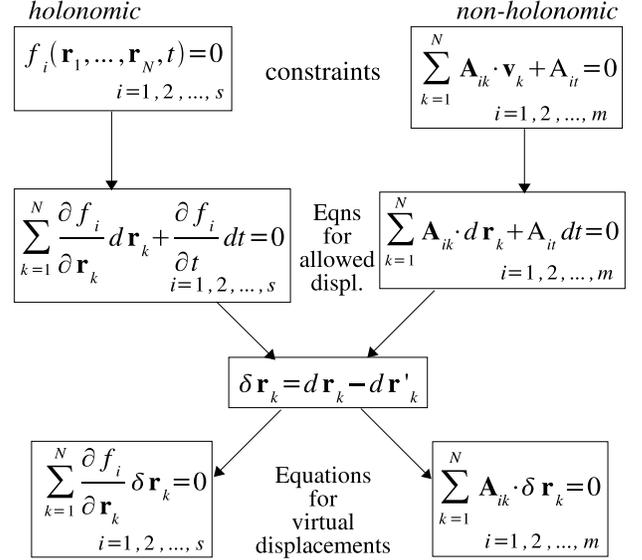}}
\caption{Virtual displacement defined as difference of allowed
displacements}
\label{connect}
\end{figure}

The virtual displacements thus defined, satisfy the homogeneous 
part of the constraint equations, i.e., Eq.(\ref{allcons1}) and
Eq.(\ref{allcons2}) with $\partial f_i/\partial t =0$ and $A_{it} = 0$.
Hence,
\beq{\label{vircons1}}
\sum_{k=1}^{N} \frac{\partial f_i}{\partial \mbf{r}_k} 
\cdot \delta\mbf{r}_k = 0, 
\hfill i=1,2,\dots,s,
\eeq
\vskip -0.5cm
\beq{\label{vircons2}}
\sum_{k=1}^{N} \mbf{A}_{ik} \cdot \delta \mbf{r}_k  = 0,
\hfill i=1,2,\dots,m.
\eeq
The logical connection between the equations of constraint,
equations for allowed displacements and equations for virtual
displacements are presented in FIG.\ref{connect}.

The absence of the $(\partial f_i/\partial t)$ and $A_{it}$
in the above equations, Eq.(\ref{vircons1}) and Eq.(\ref{vircons2}),
gives the precise meaning to the statement:
``{\it virtual displacements are the allowed
displacements in the case of frozen constraints}''.
The constraints are frozen in time
in the sense that we make the $(\partial f_i/\partial t)$ and
$A_{it}$ terms zero,
though the $\partial f_i /\partial \mbf{r}_k$ and
$\mbf{A}_{ik}$ terms still involve both position
$\{\mbf{r}_1,\mbf{r}_2,\dots,\mbf{r}_N\}$, and time ($t$).
In the case of stationary constraints, i.e.,
$f_i(\mbf{r}_1,\dots,\mbf{r}_N)=0$, and
$\sum_{k} \mbf{A}_{ik}(\mbf{r}_1,\dots,\mbf{r}_N) \cdot \mbf{v}_k =0$,
the virtual displacements are identical with
allowed displacements as $(\partial f_i/\partial t)$ and
$A_{it}$ are identically zero.
\subsection{Existence of forces of constraints}
In the case of an unconstrained system of $N$ particles
described by position vectors $\{\mbf{r}_k\}$ and  velocity
vectors $\{\mbf{v}_k\}$, the motion is governed by Newton's Law,
\beq{\label{newton}}
m_k \mbf{a}_k = \mbf{F}_k (\mbf{r}_l,\mbf{v}_l,t),
\hfill k, l =1,2,\dots,N
\eeq
where $m_k$ is the mass of the $\mathrm k^{th}$ particle, $\mbf{a}_k$ is
its acceleration and $\mbf{F}_k$ is the total external force acting
on it. However, for a constrained system, the equations of
constraint, namely Eq.(\ref{holo}) and Eq.(\ref{nonholo}),
impose the following restrictions on the allowed accelerations,
\beqar{\label{acccons1}}
\sum_{k=1}^{N} \frac{\partial f_i}{\partial \mbf{r}_k}
\cdot \mbf{a}_k + \sum_{k=1}^{N} \frac{d}{dt}\left(\frac{\partial
f_i}{\partial \mbf{r}_k}\right) \mbf{v}_k +\frac{d}{dt}\left(
\frac{\partial f_i}{\partial t}\right) = 0, \nonumber \\
\hfill i=1,2,\dots,s
\eeqar
\vskip -1.0cm
\beqar{\label{acccons2}}
\sum_{k=1}^{N} \mbf{A}_{ik} \cdot \mbf{a}_k + \frac{d}{dt} \mbf{A}_{ik} \cdot
\mbf{v}_k + \frac{d}{dt} A_{it} = 0, \nonumber \\
\hfill i=1,2,\dots,m.
\eeqar
Given $\{\mbf{r}_k\}$, $\{\mbf{v}_k\}$ one is no longer free to choose
all the accelerations $\{\mbf{a}_k\}$ independently.
Therefore in general the accelerations $\{\mbf{a}_k\}$ allowed
by Eq.(\ref{acccons1}),  Eq.(\ref{acccons2}) are incompatible with
Newton's Law, i.e., Eq.(\ref{newton}).

This implies that during the motion the constraint
condition cannot be maintained by the external forces alone.
Physically some additional forces, e.g., normal reaction from 
the surface of constraint, tension in the pendulum string,
come into play to ensure that the constraints are satisfied
throughout the motion.
Hence one is compelled to introduce forces of constraints
$\{\mbf{R}_k\}$ and modify the equations of motion as,
\beq{\label{newtonconst}}
m_k \mbf{a}_k = \mbf{F}_k + \mbf{R}_k, 
\hfill k=1,2,\dots,N.
\eeq

\begin{figure}[h]
{\includegraphics{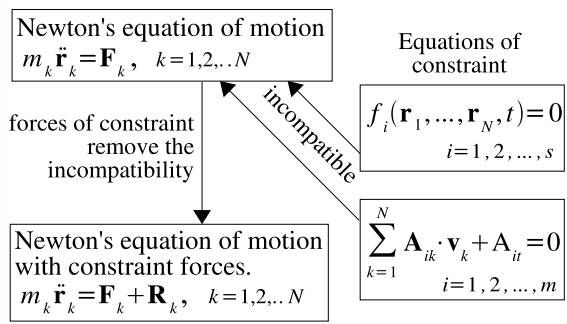}}
\caption{Existence of constraint forces}
\label{const_for}
\end{figure}

FIG.\ref{const_for} presents the connection between the
equations of constraints and forces of constraints.

Now the problem is to determine the motion of $N$ particles,
namely their positions $\{\mbf{r}_k(t)\}$, velocities $\{\mbf{v}_k(t)\}$
and the forces of constraints $\{\mbf{R}_k\}$, for a given set of
external forces $\{\mbf{F}_k\}$, constraint equations,
Eq.(\ref{holo}) and Eq.(\ref{nonholo}),
and initial conditions $\{\mbf{r}_k(0), \mbf{v}_k(0)\}$.
It is important that the initial conditions are also
compatible with the constraints.
There are a total of ($6N$) scalar unknowns, namely the
components of $\mbf{r}_k(t)$ and $\mbf{R}_k$, connected by
($3N$) scalar equations of motion, Eq.(\ref{newtonconst}), and ($s+m$)
equations of constraints, Eq.(\ref{holo}) and Eq.(\ref{nonholo}).
For ($6N > 3N + s + m$) we have an under-determined
system. Hence to solve this problem we need ($3N-s-m$) additional
scalar relations.

\subsection{Solvability and ideal constraints}
\begin{figure}[h]
{\includegraphics{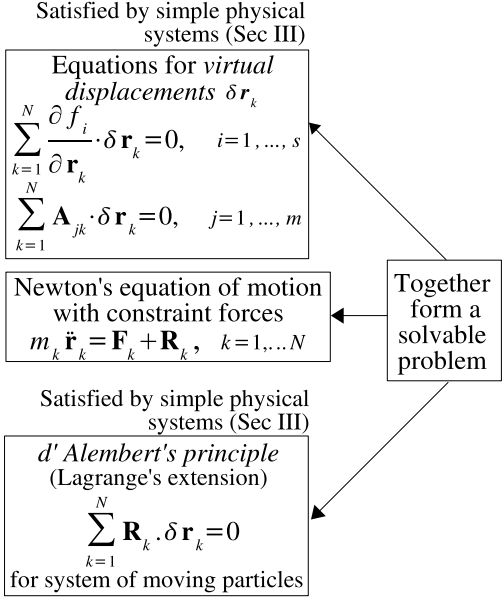}}
\caption{Solvability under ideal constraints}
\label{solvable}
\end{figure}

In simple problems with stationary constraints, e.g., motion 
of a particle on
a smooth stationary surface, we observe that the allowed
displacements are tangential to the surface. The virtual 
displacement being a difference of two such allowed displacements,
is also a vector tangential to it. 
For a frictionless surface, the force of constraint,
the so called `normal reaction', is perpendicular to the surface.
Hence the work done by the constraint force is zero, on allowed 
as well as virtual displacement.
\bd
\sum_{k=1}^N \mbf{R}_k \cdot d\mbf{r}_k = 0, 
\hskip 1cm 
\sum_{k=1}^N \mbf{R}_k \cdot \delta\mbf{r}_k = 0.
\ed
When the constraint surface is in motion, the allowed
velocities, and hence the allowed displacements are no 
longer tangent to the surface (see Sec.III). The virtual 
displacement however remains tangent to the constraint 
surface. 
As the surface is frictionless, it is natural to assume
that the force of constraint is still
normal to the instantaneous position of the surface.
Hence the work done by normal reaction on virtual displacement
is zero. However the work done by
constraint force on allowed displacements is no longer zero.
\beq{\label{vwork}}
\sum_{k=1}^N \mbf{R}_k \cdot d\mbf{r}_k \neq 0, \hskip 1cm 
\sum_{k=1}^N \mbf{R}_k \cdot \delta\mbf{r}_k = 0.
\eeq
In a number of physically interesting simple problems, such
as, motion of a pendulum with fixed or moving support, 
motion of a particle along a stationary and moving slope, 
we observe that the above interesting relation 
between the force of constraint and virtual 
displacement holds (see Sec.III). 
As the $3N$ scalar components of the virtual displacements
$\{\delta\mbf{r}_k\}$ are connected by ($s+m$) equations,
Eq.(\ref{vircons1}) and Eq.(\ref{vircons2}), only $n=(3N-s-m)$
of these scalar components are independent.
If the ($s+m$) dependent quantities are expressed 
in terms of remaining ($3N-s-m$) independent objects we get,
\beq{\label{vwork2}}
\sum_{j=1}^n \widetilde{R}_j \cdot \delta \widetilde{x}_j = 0.
\eeq
where $\{\widetilde{x}_j\}$ are the independent components of 
$\{\mbf{r}_k\}$. $\{\widetilde{R}_j\}$ are the coefficients of 
$\{\delta \widetilde{x}_j\}$, and are composed of different $\{\mbf{R}_k\}$.
Since the above components of virtual 
displacements $\{\delta \widetilde{x}_j\}$ are independent, 
one can equate each of their coefficients to zero 
($\widetilde{R}_j =0$).
This brings in exactly ($3N-s-m$) new scalar conditions or equations
that are needed to make the system solvable (see FIG.\ref{solvable}).

Thus we have found a special class of constraints,
which is observed in nature (Sec.III) and 
which gives us a solvable mechanical system.
We call this special class of constraints, 
where the forces of constraint do zero work on 
virtual displacement, i.e., 
$\sum_k \mbf{R}_k \cdot \delta\mbf{r}_k = 0$,
the {\it ideal constraint}.

Our interpretation of the principle of zero virtual work, as a 
definition of an ideal class of constraints, agrees with
Sommerfeld. In his exact words, ``{\it a general} {\bf postulate} 
{\it of mechanics: in any mechanical systems the virtual work of 
the reactions equals zero. Far be it from us to want to give a 
general proof of this postulate, rather we regard it practically 
as} {\bf definition of a
mechanical system}'' \cite{sommer}. 
(Boldface is added by the authors).
\section{Examples of virtual displacements}
\subsection{Simple Pendulum with stationary support}
\begin{figure}[h]
{\includegraphics{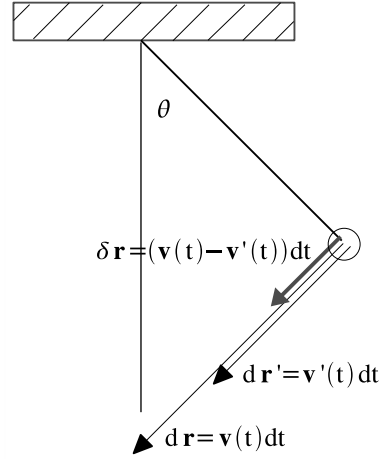}}
\caption{Allowed and virtual displacements of a pendulum with 
stationary support}
\label{pend_stat}
\end{figure}

The motion of a pendulum is confined to a plane and its bob
moves keeping a fixed distance from the point of suspension 
(see FIG.\ref{pend_stat}).
The equation of constraint therefore is,
\bd
f(x,y,t) \doteq x^2+y^2- r_0^2 = 0,
\ed
where $r_0$ is the length of the pendulum.
Whence
\bd
\frac{\partial f}{\partial x} =2x ,\hskip .5cm
\frac{\partial f}{\partial y} =2y ,\hskip .5cm
\frac{\partial f}{\partial t} = 0.
\ed

The constraint equation for allowed velocities, Eq.(\ref{velconst}),
becomes,
\bd
x \cdot v_x + y \cdot v_y =0.
\ed
Hence the allowed velocity ($v_x$, $v_y$)
is orthogonal to the instantaneous 
position ($x$,$y$) of the bob relative to stationary support.
The same may also be verified taking a plane polar coordinate.

The allowed velocities and the allowed displacements are
perpendicular to the line of suspension. 
The virtual velocities and the virtual displacements,
being the difference of two unequal allowed velocities and
displacements respectively, are
also perpendicular to the line of suspension. 
\bd
d \mbf{r} = \mbf{v}(t) dt, \hskip 1cm d \mbf{r}' = \mbf{v}'(t) dt,
\ed
\vskip -.5cm
\bd
\delta \mbf{r} = (\mbf{v}(t) - \mbf{v}'(t) ) dt.
\ed
Although the virtual displacement is not uniquely specified
by the constraint, it is restricted to be in a plane 
perpendicular to the instantaneous line of suspension.
Hence it is not `completely arbitrary'.

The ideal string of the pendulum provides a tension ($\mbf{T}$)
along its length, but no shear.
The work done by this tension on both allowed and virtual
displacements is zero,
\bd
\mbf{T} \cdot d\mbf{r} = 0, \hskip 1cm \mbf{T} \cdot \delta\mbf{r} = 0.
\ed
\subsection{Simple Pendulum with moving support}
\begin{figure}[h]
{\includegraphics{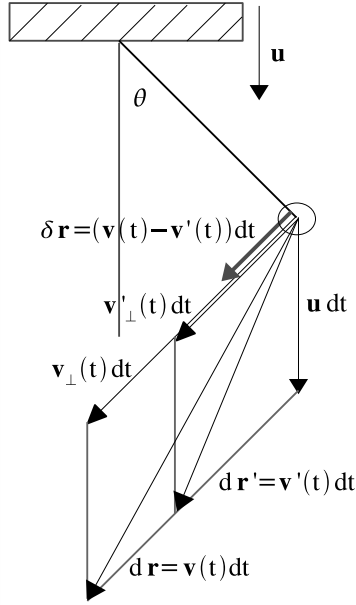}}
\caption{Allowed and virtual displacements of a pendulum with moving support}
\label{pend_mv}
\end{figure}

Let us first consider the case when the support is moving
vertically with a velocity $u$.
The motion of the pendulum is still confined to a plane.
The bob moves keeping a fixed distance from
the moving point of suspension (FIG.\ref{pend_mv}).
The equation of constraint is,
\bd
f(x,y,t) \doteq x^2+(y-u t)^2- r_0^2 = 0,
\ed
where $u$ is the velocity of the point of suspension
along a vertical direction.

Whence
\bd
\frac{\partial f}{\partial x} =2x ,\hskip .3cm
\frac{\partial f}{\partial y} =2 (y- u t) ,\hskip .3cm
\frac{\partial f}{\partial t} = - 2 u (y - ut).
\ed
Hence the constraint equation gives,
\bd
x \cdot v_x + (y- u t) \cdot v_y - u (y - ut)=0,
\ed
or,
\bd
x \cdot v_x + (y- u t) \cdot (v_y - u)  =0.
\ed
The allowed velocities ($v_x$, $v_y$)
and the allowed displacements,
are not orthogonal to the 
instantaneous position of the bob relative to
the instantaneous point of suspension ($x$, $y-ut$).
It is easy to verify from the above equation that the 
allowed velocity ($v_x$, $v_y$) is equal to 
the sum of a velocity vector ($v_x$, $v_y - u$)
perpendicular to the position of the bob relative
to the point of suspension ($x$,$y-ut$), and the 
velocity of the support ($0$, $u$).
If we denote $\mbf{v}(t)= (v_x, v_y)$, 
$\mbf{v}_{\bot}(t)= (v_x, v_y - u)$ and $\mbf{u}=(0,u)$,
then,
\bd
\mbf{v}(t) = \mbf{v}_{\bot}(t) + \mbf{u}.
\ed
The allowed displacements are vectors collinear to allowed 
velocities. A virtual displacement being the difference of 
two allowed displacements, is a vector collinear to the
difference of allowed velocities. Hence it is orthogonal
to the instantaneous line of suspension.
\beqa
d \mbf{r}&=&\mbf{v}(t) dt = \mbf{v}_{\bot}(t) dt + \mbf{u} dt, \\
d \mbf{r}'&=&\mbf{v}'(t) dt = \mbf{v}_{\bot}'(t) dt + \mbf{u} dt, \\
\delta \mbf{r}&=&(\mbf{v}(t) - \mbf{v}'(t) ) dt = 
(\mbf{v}_{\bot}(t) -\mbf{v}'_{\bot}(t) ) dt.
\eeqa
Hence none of these allowed or virtual vectors are `arbitrary'.

At any given instant, an ideal string provides a tension along its length,
with no shear. Hence the constraint force, namely tension $\mbf{T}$,
does zero work on virtual displacement.
\bd
\mbf{T} \cdot d\mbf{r} = \mbf{T} \cdot \mbf{u} \; dt \neq 0, 
\hskip 1cm \mbf{T} \cdot \delta\mbf{r} = \mbf{T} \cdot \mbf{v}_{\bot} \; dt = 0
\ed
For the support moving in a horizontal or any arbitrary direction,
one can show that the allowed displacement is not normal to the 
instantaneous line of suspension. 
But the virtual displacement, as defined in this article,
always remains perpendicular to the instantaneous line of support.
\subsection{Motion along a stationary inclined plane}
\begin{figure}[h]
{\includegraphics{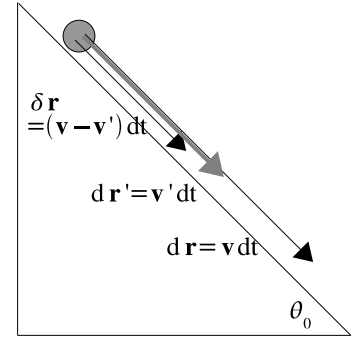}}
\caption{Allowed and virtual displacements of a particle 
sliding along a stationary slope}
\label{slop_stat}
\end{figure}

Let us consider a particle sliding along a stationary inclined plane
as shown in FIG.\ref{slop_stat}.
The constraint here is more conveniently expressed in polar
coordinates. The constraint equation is,
\bd
f(r,\theta) \doteq \theta - \theta_0 =0
\ed
where $\theta_0$ is the angle of the slope. 
Hence the constraint equation
for allowed velocities, Eq.(\ref{velconst}), gives,
\beqa
\left( \frac{\partial f}{\partial \mbf{r}} 
\right) \cdot \mbf{v} + \frac{\partial f}{\partial t} &=& 
\left( \frac{\partial f}{\partial r} \right) v_r +
\left( \frac{\partial f}{\partial \theta} \right) v_{\theta} +
\frac{\partial f}{\partial t} \\
&=& 0 \cdot \dot{r} + 1 \cdot (r \dot{\theta}) + 0 = 0
\eeqa
Hence $\dot{\theta}=0$, implying that the allowed velocities are 
along the constant $\theta$ plane. Allowed velocity,
allowed and virtual displacements are,
\beqa
&& \mbf{v} = \dot{r} \,\widehat{\mbf{r}}, \hskip .5cm
d\mbf{r} = \dot{r} \,\widehat{\mbf{r}} \,dt, \nonumber \\
&& \mbf{v}' = \dot{r}' \,\widehat{\mbf{r}}, \hskip .5cm
d\mbf{r}' = \dot{r}' \,\widehat{\mbf{r}} \,dt, \nonumber \\
&& \delta\mbf{r} = (\mbf{v} - \mbf{v}') dt =
(\dot{r}-\dot{r}')\, \widehat{\mbf{r}} \,dt.
\eeqa
where $\widehat{\mbf{r}}$ is a unit vector along the slope.

As the inclined slope is frictionless, the constraint
force $\mbf{N}$ is normal to the surface.
The work done by this force on allowed as well as virtual
displacement is zero.
\bd
\mbf{N} \cdot d\mbf{r} = 0, \hskip 1cm \mbf{N} \cdot \delta\mbf{r} = 0
\ed
\subsection{Motion along a moving inclined plane}
\begin{figure}[h]
{\includegraphics{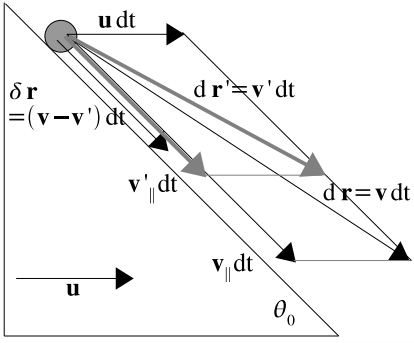}}
\caption{Allowed and virtual displacements of a particle 
sliding along a moving slope}
\label{slop_mv}
\end{figure}

For an inclined plane moving along the horizontal side 
(FIG.\ref{slop_mv}), the constraint is given by,
\beqa
\frac{(x- u t)}{y} - \cot(\theta_0) &=& 0, \\
f(x,y) \doteq (x - u t) - \cot(\theta_0) y &=& 0.
\eeqa
Whence the constraint for allowed velocities Eq.(\ref{velconst})
becomes,
\bd
(\dot{x} - u) - \cot(\theta_0) \dot{y} = 0.
\ed

Hence the allowed velocity ($\dot{x},\dot{y}$) is the sum of
two vectors, one along the plane ($\dot{x} - u,\dot{y}$), and
the other equal to the velocity of the plane itself ($u,0$).
\bd
\mbf{v}(t) = \mbf{v}_{\|}(t) + \mbf{u}.
\ed
Allowed displacements are vectors along the allowed velocities,
however the virtual displacement is still a vector along the
instantaneous position of the plane.
\beqa
d \mbf{r}&=&(\mbf{v}_{\|}(t)+ \mbf{u})dt, \;\;\;\;\;\; 
d\mbf{r}'=(\mbf{v}'_{\|}(t) + \mbf{u}) dt, \\
\delta \mbf{r}&=&(\mbf{v}(t)-\mbf{v}'(t)) dt = 
(\mbf{v}_{\|}(t)-\mbf{v}'_{\|}(t)) dt.
\eeqa
For the moving frictionless slope, the constraint
force provided by the surface is perpendicular to the plane.
Hence the work done by the constraint force on virtual
displacement remains zero.
\bd
\mbf{N} \cdot d\mbf{r} \neq 0, \hskip 1cm \mbf{N} \cdot \delta\mbf{r} = 0.
\ed
\section{Lagrange's method of undetermined multipliers}
A constrained system of particles obey the equations of
motion given by,
\bd
m_k \mbf{a}_k = \mbf{F}_k + \mbf{R}_k,
\hfill k=1,2,\dots,N
\ed
where $m_k$ is the mass of the $\mathrm{k^{th}}$ particle,
$\mbf{a}_k$ is its acceleration. $\mbf{F}_k$
and $\mbf{R}_k$ are the total external force and force
of constraint on the $\mathrm{k^{th}}$ particle.
If the constraints are {\it ideal}, we can write,
\beq{\label{vwork00}}
\sum_{k=1}^N \mbf{R}_k \cdot \delta\mbf{r}_k = 0,
\eeq
whence we obtain,
\beq{\label{geneqn}}
\sum_{k=1}^N (m_k \mbf{a}_k - \mbf{F}_k) \cdot 
\delta \mbf{r}_k = 0.
\eeq
If the components of $\{\delta \mbf{r}_k\}$ were independent, we
could recover Newton's Law for unconstrained system from this
equation. However for a constrained system $\{\delta \mbf{r}_k\}$ 
are dependent through the constraint equations,
Eq.(\ref{vircons1}) and Eq.(\ref{vircons2}), for
holonomic and non-holonomic systems respectively.
\setcounter{equation}{13}
\beqar
&& \sum_{k=1}^N \frac{\partial f_i}{\partial \mbf{r}_k} \; \delta \mbf{r}_k = 0,
\hskip 1cm i=1,2,\dots,s \\
&& \sum_{k=1}^{N} \mbf{A}_{jk} \cdot \delta \mbf{r}_k  = 0,
\hskip 1cm j=1,2,\dots,m
\eeqar
\setcounter{equation}{23}
We multiply Eq.(\ref{vircons1}) successively by ($s$) scalar multipliers 
$\{\lambda_1, \lambda_2, \dots \lambda_s\}$, 
Eq.(\ref{vircons2}) successively by ($m$) scalar multipliers
$\{\mu_1, \mu_2, \dots \mu_m\}$
and then subtract them from the zero virtual work equation, 
namely Eq.(\ref{vwork00}).
\beq{\label{vwork3}}
\sum_{k=1}^N \left(\mbf{R}_k - 
\sum_{i=1}^s \lambda_i \frac{\partial f_i}{\partial \mbf{r}_k} - 
\sum_{j=1}^m \mu_j \mbf{A}_{jk} 
\right) 
\delta \mbf{r}_k = 0.
\eeq
These multipliers $\{\lambda_i\}$ and $\{\mu_j\}$ are
called the Lagrange's multipliers.
Explicitly in terms of components,
\beqar{\label{vwork4}}
\sum_{k=1}^N \left[ R_{k,x} - 
\sum_{i=1}^s \lambda_i \frac{\partial f_i}{\partial x_k} -
\sum_{j=1}^m \mu_j (\mbf{A}_{jk})_{_x}  
\right] \delta x_k \nonumber \\
+ \sum_{k=1}^N [Y]_k \delta y_k + \sum_{k=1}^N [Z]_k \delta z_k  = 0,
\eeqar
where $[Y]_k$ and $[Z]_k$ denote the coefficients of
$\delta y_k$ and $\delta z_k$ respectively.

\begin{figure}[h]
{\includegraphics{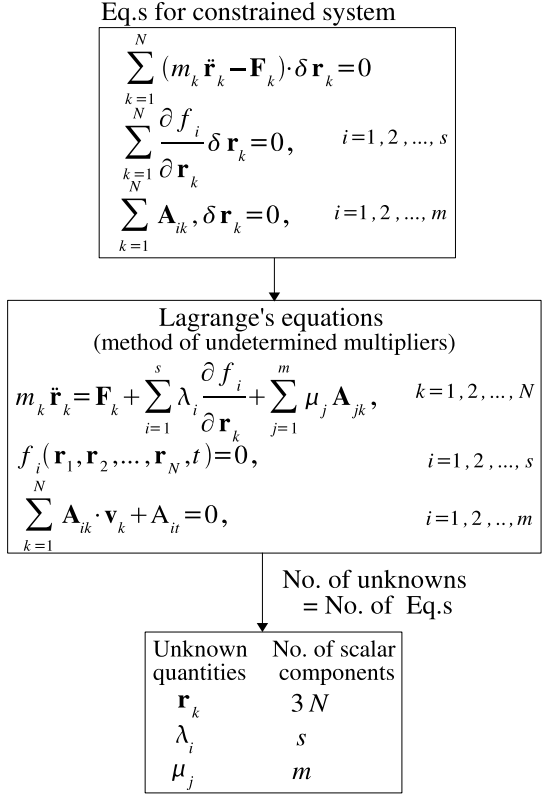}}
\caption{Lagrange's method of undetermined multipliers: solvability}
\label{un_mult}
\end{figure}

The constraint equations, Eq.(\ref{vircons1}) and Eq.(\ref{vircons2}),
allow us to 
write the ($s+m$) dependent virtual displacements in terms of 
the remaining $n=(3N-s-m)$ independent ones. We choose ($s+m$) 
multipliers $\{\lambda_1,\lambda_2,\dots,\lambda_s\}$
and $\{\mu_1, \mu_2,\dots, \mu_m \}$,
such that the coefficients of ($s+m$) dependent
components of virtual displacement vanish. The remaining
virtual displacements being independent, their coefficients 
must vanish as well. Thus it is possible to choose 
$\{\lambda_1,\lambda_2,\dots,\lambda_s \}$
and $\{\mu_1, \mu_2,\dots, \mu_m \}$ such that all coefficients 
$\{[X]_k$,$[Y]_k$,$[Z]_k \}$
of virtual displacements 
$\{\delta x_k$,$\delta y_k$,$\delta z_k \}$ in
Eq.(\ref{vwork4}) vanish. Hence we can express the forces of constraint 
in terms of the Lagrange's multipliers.
\beq{\label{fconst}}
\mbf{R}_k = \sum_{i=1}^s \lambda_i \frac{\partial f_i}{\partial \mbf{r}_k} +
\sum_{j=1}^m \mu_j \mbf{A}_{jk}, 
\hfill k =1,2,\dots, N
\eeq
Thus the problem reduces to finding a solution for the equations,
\beq{\label{leq_undet}}
m_k \mbf{a}_k = \mbf{F}_k + \sum_{i=1}^s \lambda_i 
\frac{\partial f_i}{\partial \mbf{r}_k} +
\sum_{j=1}^m \mu_j \mbf{A}_{jk}, 
\hfill k=1,2,\dots,N
\eeq
together with the equations of constraint,
\setcounter{equation}{5}
\beq
f_i(\mbf{r}_1,\mbf{r}_2,\dots,\mbf{r}_N,t) = 0,
\hfill i=1,2,\dots,s
\eeq
and
\beq
\sum_{k=1}^{N} \mbf{A}_{ik} \cdot \mbf{v}_k + 
A_{it} = 0\,, 
\hfill i=1,2,\dots,m.
\eeq
\setcounter{equation}{27}
Here we have to solve ($3N+s+m$) scalar equations involving
($3N+s+m$) unknown scalar quantities, namely $\{x_k, y_k, z_k,
\lambda_i, \mu_j\}$ (see FIG.\ref{un_mult}).
After solving this system of equations for
$\{x_k, y_k, z_k, \lambda_i, \mu_j\}$, one can obtain the
forces of constraint $\{\mbf{R}_k\}$ using Eq.(\ref{fconst}).
\section{Lagrange's Equations in Generalized Coordinates}
\begin{figure}[h]
{\includegraphics{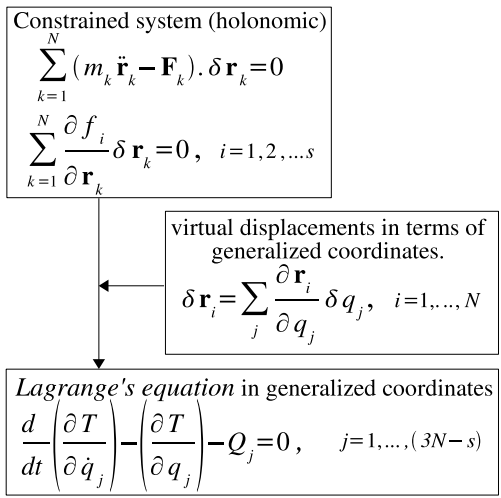}}
\caption{Lagrange's equations in generalized coordinates}
\label{gen_co}
\end{figure}

For the sake of completeness we discuss very briefly Lagrange's 
equations in generalized coordinates.
A more complete discussion can be found in most texts
\cite{goldstein,pars,sommer,hylleraas,greenwood,schaum,symon,sygr,
taylor,haas,terhaar,hand,landau,arnold}.
Consider a system of $N$ particles under ($s$) holonomic
and ($m$) non-holonomic constraints.
In certain suitable cases, one can express ($s+m$) dependent coordinates
in terms of the remaining ($3N-s-m$) independent ones.
It may be noted that such a complete reduction is not possible for general
cases of non-holonomic and time dependent constraints \cite{sommer,pars}.
If we restrict our discussion to cases where this reduction is
possible, one may express all the $3N$ scalar components of position
$\{\mbf{r}_1, \mbf{r}_2, \dots, \mbf{r}_N\}$
in terms of ($3N-s-m$) independent parameters $\{q_1, q_2,\dots,q_n\}$
and time ($t$).
\beq
\mbf{r}_k = \mbf{r}_k(q_1, q_2, \dots, q_n, t),
\hfill k=1,2,\dots,N
\eeq
The allowed and virtual displacements are given by,
\beqar
d \mbf{r}_k&=&\sum_{j=1}^n \frac{\partial \mbf{r}_k}
{\partial q_j} \delta q_j+ \frac{\partial \mbf{r}_k}{\partial t}dt,
\hskip 1.2cm k=1,2,\dots,N \nonumber \\ 
\delta \mbf{r}_k&=&\sum_{j=1}^n \frac{\partial \mbf{r}_k}
{\partial q_j} \delta q_j,
\hskip 2.5cm  k=1,2,\dots,N. 
\eeqar
From Eq.(\ref{geneqn}) we obtain,
\beq\label{leqn1}
\sum_{k=1}^N m_k \frac{d\dot{\mbf{r}}_k}{dt} \left(\sum_{j=1}^n 
\frac{\partial\mbf{r}_k}{\partial q_j} \delta q_j \right) 
- \sum_{k=1}^N \mbf{F}_k
\left( \sum_{j=1}^n \frac{\partial\mbf{r}_k}{\partial q_j} 
\delta q_j \right) = 0
\eeq
Introduce the expression of kinetic energy,
\bd
T = \frac{1}{2} \sum_{k=1}^N m_k \dot{\mbf{r}}^2_k,
\ed
and that of the generalized force,
\beq
Q_j = \sum_{k=1}^N \mbf{F}_k \frac{\partial\mbf{r}_k}{\partial q_j},
\hfill j=1,2,\dots,n.
\eeq
After some simple algebra one finds,
\beq\label{leqn2}
\sum_{j=1}^n \left(\frac{d}{dt}\frac{\partial T}{\partial \dot{q}_j}
- \frac{\partial T}{\partial q_j} - Q_j \right) \delta q_j = 0.
\eeq
As $\{q_1,q_2,\dots,q_n\}$ are independent coordinates, coefficient of each 
$\delta q_j$ must be zero separately. Hence (see FIG.\ref{gen_co}),
\beq{\label{leqnGF}}
\frac{d}{dt}\frac{\partial T}{\partial \dot{q}_j}
- \frac{\partial T}{\partial q_j} = Q_j,
\hfill j=1,2,\dots,n.
\eeq
These are called Lagrange's equations in generalized coordinates.
To proceed further one has to impose additional conditions on the
nature of forces $\{\mbf{F}_k\}$ or $\{Q_j\}$.

\vfill

In problems where forces $\{\mbf{F}_k\}$ are derivable from a 
scalar potential $\widetilde{V}(\mbf{r}_1,\mbf{r}_2,\dots,\mbf{r}_N)$,
\beq
\mbf{F}_k = - \mbox{\boldmath $\nabla$}_k \widetilde{V}(\mbf{r}_1,\mbf{r}_2,
\dots,\mbf{r}_N),
\hfill k=1,2,\dots,N.
\eeq
One can obtain the generalized force as,
\beq
Q_j = - \mbox{\boldmath $\nabla$}_k \widetilde{V} \cdot
\left( \frac{\partial{\mbf{r}_k}}{\partial q_j} \right)
= - \frac{\partial V}{\partial q_j},
\hfill j=1,2,\dots,n.
\eeq
Where $V$ is the potential $\widetilde{V}$ expressed as a function of 
$\{q_1, q_2, \dots, q_n\}$.
In addition as the potential $V$ is independent of
the generalized velocities, we obtain from Eq.(\ref{leqnGF}),
\beq{\label{leqn3}}
\frac{d}{dt} \frac{\partial(T-V)}{\partial \dot{q}_j} - 
\frac{\partial(T-V)}{\partial q_j} = 0, 
\hfill j=1,2,\dots,n.
\eeq
At this stage one introduces the Lagrangian function, $L = T -V$.
In terms of the Lagrangian, the equations of motion
take up the form,
\beq{\label{leqn4}}
\frac{d}{dt}\frac{\partial L}{\partial \dot{q}_j} -
\frac{\partial L}{\partial q_j} = 0, 
\hfill j=1,2,\dots,n.
\eeq
\section{Conclusion}
In this article we make an attempt to present a
quantitative definition of the virtual displacement.
We show that for certain simple cases the work done
by the forces of constraint on virtual displacement is zero.
We also demonstrate that this zero virtual work principle gives us
a solvable class of problems. Hence we define 
this special class of constraint as {\it the ideal constraint}.
We demonstrate in brief how one can solve a general mechanical 
problem by: i) Lagrange's method of undetermined multipliers and 
ii) Lagrange's equations in generalized coordinates.

In the usual presentations of Lagrange's equation based on
virtual displacement and d'Alembert's principle, 
Eq.(3) and Eq.(5) are satisfied by the virtual displacements.
One may consider these equations as the definition of
virtual displacements.
However the situation is far from satisfactory, as 
separate defining equations are required for different 
classes of constraints.
This adhoc definition also fails to clarify the actual
connection between the virtual displacements and the equations
of constraints.

At this stage one introduces d'Alembert's principle
of zero virtual work. Bernoulli \cite{sommer} (1717) 
and d'Alembert \cite{sommer,dalembert} (1743) 
originally proposed this principle for a system in static equilibrium.
The principle states 
that the forces of constraint do zero work on virtual displacement.
For systems in static equilibrium,
virtual displacement meant an imaginary displacement
of the system that keeps its statical equilibrium unchanged.
Lagrange generalized this principle to a constrained system of 
particles in motion.
This principle is crucial in arriving at Lagrange's equation. 
However, most texts do not clearly address the questions,
(i) why one needs to extend d'Alembert's principle to particles
in motion, and
(ii) why the work done by constraint forces on virtual displacements,
and not on allowed displacements, is zero ?

In the present article the allowed infinitesimal displacements
are defined as ones that satisfy the infinitesimal form of 
the constraint equations. They are the displacements that could have
been possible if only the constraints were present. Actual dynamics,
under the given external forces, would choose one of these
various sets of displacements as actual displacement of the system. 
The definition of virtual displacement
as difference of two unequal allowed displacements over the same 
infinitesimal time interval ($t, t+dt$), gives a unified
definition of virtual displacement. 
This definition of virtual displacement satisfies
the appropriate equations found in the literature, for both
holonomic and non-holonomic systems.

It is shown that Newton's equation of motion with external forces alone,
is inconsistent with equations of constraint. 
Hence the forces of constraint are introduced. 
Now there are $3N$ equations of motion and $(s+m)$ equations of constraint
involving $6N$ unknown scalars $\{ \mbf{r}_k(t), \mbf{R}_k \}$.
Without additional condition (d'Alembert principle),
the problem is underspecified and unsolvable. 

It is verified that for simple physical systems,
the virtual displacements, as defined in this article,
satisfy d'Alembert principle 
for particles in motion. The rheonomous examples discussed in Sec.III
show why the forces of constraint do zero work on virtual displacements
and not on allowed displacements.
The additional equations introduced by d'Alembert principle
make the problem solvable.
These justify
(i) the {\it peculiar} definition of virtual displacements and the
equations they satisfy,
(ii) Lagrange's extension of d'Alembert's principle to particles
in motion,
(iii) why the zero work principle is related to virtual displacement
and not to allowed displacement.
Once the system is solvable, two methods, originally proposed 
by Lagrange can be used, and are demonstrated.
For Lagrange's method of undetermined multipliers, one
solves ($3N+s$) equations to obtain the motion of the system
$\{\mbf{r}_k(t),\dots,\mbf{r}_N(t)\}$ 
and Lagrange's multipliers $\{\lambda_i,\dots,\lambda_s \}$.
The forces of constraint $\{\mbf{R}_k,\dots,\mbf{R}_N \}$
are expressed in terms of
these multipliers.
For Lagrange's equations in generalized coordinates one
solves ($3N-s$) equations to obtain the time evolution of
the generalized coordinates $\{q_j=q_j(t), \; j=1, \dots,3N-s\}$.
This gives the complete description of the motion
$\{\mbf{r}_k=\mbf{r}_k(q_1,q_2,\dots,q_n,t), \; k=1,\dots,N\}$, 
ignoring the calculation of the constraint forces.
It may be noted that about a century later Appell's
equations \cite{sommer,pars,appell}
were introduced for efficiently solving non-holonomic systems.

\begin{figure*}
{\includegraphics{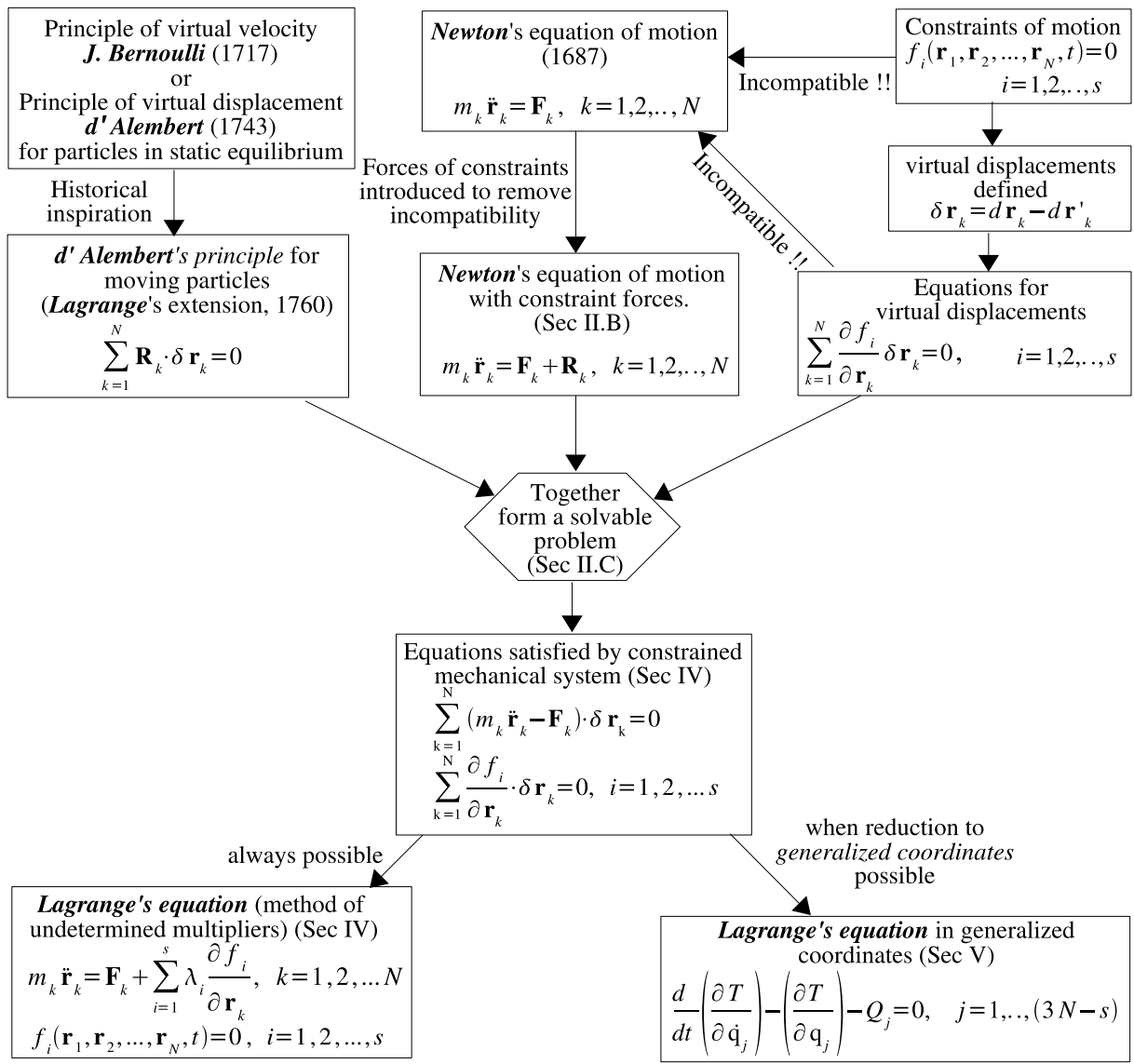}}
\caption{Logical connection between constraint equations,
virtual displacements, forces of constraints, d' Alembert's principle
and Lagrange's equations.}
\label{logical}
\end{figure*}

It is interesting to note that both the above mentioned
methods require the principle of zero virtual work by
constraint forces as a crucial starting point.
In the case of Lagrange's method of undetermined
multipliers we start with the ideal constraint condition
Eq.(\ref{vwork00}). From there we obtain
Eq.(\ref{geneqn})-Eq.(\ref{leq_undet}).
Eq.(\ref{fconst}) expresses the constraint forces in terms
of Lagrange's multipliers.
For Lagrange's equations in generalized coordinates,
we start with the ideal constraint, Eq.(\ref{vwork00}).
We work our way through Eq.(\ref{geneqn}), Eq.(\ref{leqn1}),
Eq.(\ref{leqn2}) and finally obtain Lagrange's
equations in generalized coordinates, Eq.(\ref{leqnGF})
and Eq.(\ref{leqn4}). The last figure, FIG.(\ref{logical}),
gives the complete logical flow of this article.
\section*{Acknowledgment}
Authors wish to thank
Professor John D. Jackson and Professor Leon
A. Takhtajan for their remarks and suggestions.
They also thank Professor Sidney Drell,
Professor Donald T. Greenwood and Professor Alfred
S. Goldhaber for their valuable communications.

The authors gratefully acknowledge the encouragement received
from Professor Max Dresden at Stony Brook.
Authors have greatly benefited from the books mentioned in this
article, particularly those of Arnold \cite{arnold}, 
Goldstein \cite{goldstein}, Greenwood \cite{greenwood}, 
Pars \cite{pars} and Sommerfeld \cite{sommer}.

The material presented in this article was used as part of Classical
Mechanics courses at Jadavpur University during 2000-2003.
SR would like to thank his students A. Chakraborty and B. Mal
for meaningful discussions.

\end{document}